\documentclass[aps,12pt,preprintnumbers,nofootinbib,superscriptaddress]{revtex4}
\usepackage{graphicx}
\usepackage{amssymb,amsmath}

\begin{document}

\title{New Phase Transition Related to the Black Hole's Topological Charge}

\author{Shan-Quan Lan}
\email[ ]{shanquanlan@126.com}
\affiliation{%
Department of Physics, Lingnan Normal University, Zhanjiang, 524048, Guangdong, China}
\author{Gu-Qiang Li}
\affiliation{%
Department of Physics, Lingnan Normal University, Zhanjiang, 524048, Guangdong, China}
\author{Jie-Xiong Mo }
\affiliation{%
Department of Physics, Lingnan Normal University, Zhanjiang, 524048, Guangdong, China}
\author{Xiao-Bao Xu }
\affiliation{%
Department of Physics, Lingnan Normal University, Zhanjiang, 524048, Guangdong, China}

\begin{abstract}
   The topological charge $\epsilon$ of AdS black hole is introduced by Tian etc in their papers, where a complete thermodynamic first law is obtained. In this paper, we investigate a new phase transition related to the topological charge in Einstein-Maxwell theory. Firstly, we derive the explicit solutions corresponding to the divergence of specific heat $C_{\epsilon}$ and determine the phase transition critical point. Secondly, the $T-r$ curve and $T-S$ curve are investigated and they exhibit an interesting van der Waals system's behavior. Critical physical quantities are also obtained which are consistent with those derived from the specific heat analysis. Thirdly, a van der Waals system's swallow tail behavior is observed when $\epsilon>\epsilon_{c}$ in the $F-T$ graph. What's more, the analytic phase transition coexistence lines are obtained by using the Maxwell equal area law and free energy analysis, the results of which are consistent with each other.
\end{abstract}
\maketitle
\newpage

\date{\today}

\section{Introduction }
\label{intro}

Black hole is a complicated object, there are Hawking radiation, entropy and phase transition, etc.  Although black hole's microscopic mechanism is still not clear, its thermodynamic properties can be systematically studied as it is a thermodynamic system which are described by only few physical quantities, such as mass, charge, angular momentum, temperature, entropy, etc. Generally, these thermodynamic quantities are described on the horizon and they are related by the first law. However, they can be generalised on surface out of the horizon\cite{Chen:2010ay,Brown:1992br,Brown:1992bq}. This has gotten new attention with the development of AdS/CFT, since the black hole thermodynamics on holographic screen has acquired a new and interesting interpretation as a duality of the correspondence field theory\cite{Bredberg:2011jq}.

In Ref.\cite{Tian:2014goa,Tian2018hlw}, a maximally symmetric black hole thermodynamics on holographic screen are studied in Einstein-Maxwell's gravity and Lovelock-Maxwell theory. The author found a topological charge naturally arisen in holography.  Together with all other known charges ( electric charge, mass, entropy\cite{Wald:1993nt}), they satisfy an extended first law and the Gibbs-Duhem-like relation as a completeness.  Based on the extended first law in Einstein-Maxwell's gravity, we will investigate the black hole's possible phase transition phenomenon related to the topological charge.

This paper is organized as follows. In Sec.\ref{setup} we will briefly review how the extended first law is obtained in Ref.\cite{Tian2018hlw}. In Sec.\ref{newpt}, by analysing the specific heat, the phase transition of AdS black hole in 4 dimensional space-time is studied and the critical point is determined. Then the van der Waals like behavior of temperature are observed in both $T-r$ graph and $T-S$ graph in Sec.\ref{tofrs}. In  Sec.\ref{maxfreephase}, we use the Maxwell equal area law and free energy to have obtained a consistent phase transition coexistence line.  Conclusions is drawn in Sec.\ref{con}.

\section{Review of the Topologically Charged AdS Black Holes}
\label{setup}
A d dimensional space-time AdS black hole solution with the extra topological charge in the Einstein-Maxwell theory was investigated in Ref.\cite{Tian:2014goa,Tian2018hlw}. The metric reads
\begin{equation}\label{metric}
  ds^{2}=\frac{dr^{2}}{f(r)}-f(r)dt^{2}+r^{2}d\Omega^{(k)2}_{d-2},
\end{equation}
where
\begin{eqnarray}
  f(r)&=&k+\frac{r^{2}}{l^{2}}-\frac{m}{r^{d-3}}+\frac{q^{2}}{r^{2d-6}},\nonumber\\
  d\Omega^{(k)2}_{d-2}&=&\hat{g}^{(k)}_{ij}(x)dx^{i}dx^{j},\nonumber\\
  A&=&-\frac{\sqrt{d-2}q}{\sqrt{2(d-3)}r^{d-3}}dt.
\end{eqnarray}
$m,q,l$ are related to the ADM mass $M$, electric charge $Q$, and cosmological constant $\Lambda$ by
\begin{eqnarray}
  M&=&\frac{(d-2)\Omega^{(k)}_{d-2}}{16\pi}m,\nonumber\\
  Q&=&\sqrt{2(d-2)(d-3)}(\frac{\Omega^{(k)}_{d-2}}{8\pi})q,\nonumber\\
  \Lambda &=&-\frac{(d-1)(d-2)}{2l^{2}},
\end{eqnarray}
 and $\Omega^{(k)}_{d-2}$ is the volume of the ``unit" sphere, plane or hyperbola, $k$ stands for the spatial curvature of the black hole. Under suitable compactifications for $k\leq 0$, we assume that the volume of the unit space is a constant $\Omega_{d-2}=\Omega^{(k=1)}_{d-2}$ hereafter\cite{Tian:2014goa,Tian2018hlw}.

 Following Ref.\cite{Tian2018hlw}, the first law can be obtained. Considering an equipotential surface $f(r)=c$ with fixed $c$, which can be rewritten as
 \begin{equation}\label{frc}
   f(r,k,m,q)-c=k+\frac{r^{2}}{l^{2}}-\frac{m}{r^{d-3}}+\frac{q^{2}}{r^{2d-6}}-c,
 \end{equation}
defining $K\equiv k-c$, we have
\begin{eqnarray}
  df(r,k,m,q)=\frac{\partial f}{\partial r}dr+\frac{\partial f}{\partial K}dK+\frac{\partial f}{\partial m}dm+\frac{\partial f}{\partial q}dq=0.
\end{eqnarray}
Noting
\begin{eqnarray}
  &\,&\partial_{r}f=4\pi T,\,\,\,\,\,\partial_{K}f=1,\nonumber\\
  &\,&\partial_{m}f=-\frac{1}{r^{d-3}},\,\,\,\,\,\partial_{q}f=\frac{2q}{r^{2d-6}},
\end{eqnarray}
we obtain
\begin{equation}
  dm=\frac{4\pi T}{d-2}dr^{d-2}+r^{d-3}dK+\frac{2q}{r^{d-3}}dq.
\end{equation}
Multiplying both sides with an constant factor $\frac{(d-2)\Omega_{d-2}}{16\pi}$, the above equation becomes
\begin{equation}
  dM=TdS+\frac{(d-2)\Omega_{d-2}}{16\pi}r^{d-3}dK+\Phi dQ,
\end{equation}
where $T=\frac{\partial_{r} f}{4\pi}$ is the Unruh-Verlinde temperature\cite{Chen:2010ay,Tian:2010gn}, $S=\frac{\Omega_{d-2}}{4}r^{d-2}$ is the Wald-Padmanabhan entropy\cite{Wald:1993nt,Padmanabhan:2010xh}, $\Phi=\sqrt{\frac{d-2}{2(d-3)}}\frac{q}{r^{d-3}}$ is the electric potential. If we introduce a new ``charge" as in Ref.\cite{Tian:2014goa,Tian2018hlw}
\begin{equation}
  \epsilon=\Omega_{d-2}K^{\frac{d-2}{2}},
\end{equation}
and denote its conjugate potential as $\omega=\frac{1}{8\pi}K^{\frac{4-d}{2}}r^{d-3}$, then the generalized first law is
\begin{equation}
  dM=TdS+\omega d\epsilon+\Phi dQ.
\end{equation}

\section{A New Phase Transition of AdS Black Hole }
\label{newpt}

From the generalized first law, we see there is a topological charge $\epsilon$. In this section, we will investigate the phase transition of AdS black hole in $d=4$ dimensional space-time in canonical ensemble related to the topological charge rather than the electric charge. To do so, one can observe the behavior of the specific heat at constant topological charge\cite{Mo:2016apo}.

The Unruh-Verlinde temperature is
\begin{eqnarray}
  T=\frac{f'(r)}{4\pi}&=&\frac{1}{4\pi r}(K-\frac{q^{2}}{r^{2}}+\frac{3r^{2}}{l^{2}})\nonumber\\
  &=&\frac{S \epsilon-4\pi^{2}Q^{2}+12S^{2}/l^{2}}{8\pi\Omega_{2}^{1/2}S^{3/2}}
\end{eqnarray}
Setting $l=1,Q=1,\Omega_{2}=4\pi$ hereafter, the corresponding specific heat with topological charge $\epsilon$ fixed can be calculated as
\begin{eqnarray}
  C_{\epsilon}=T(\frac{\partial S}{\partial T})_{\epsilon}&=&\frac{24S^{3}+2\epsilon S^{2}-8\pi^{2}S}{12S^{2}-\epsilon S+12\pi^{2}}\nonumber\\
  &=&\frac{2\pi r^{2}(12\pi r^{4}+\epsilon r^{2}-4\pi)}{12\pi r^{4}-\epsilon r^{2}+12\pi}
\end{eqnarray}
From the denominator, we can conclude\\
(1) when $\epsilon>24\pi$, $C_{\epsilon}$ has two diverge points at
\begin{eqnarray}
  S_{\pm}=\frac{\epsilon \pm \sqrt{\epsilon^{2}-(24\pi)^{2}}}{24},
\end{eqnarray}
which corresponds to
\begin{eqnarray}
  r_{\pm}=\sqrt{\frac{\epsilon \pm \sqrt{\epsilon^{2}-(24\pi)^{2}}}{24\pi}}.
\end{eqnarray}
(2)when $\epsilon=\epsilon_{c}=24\pi$, $C_{\epsilon}$ has only one diverge point at
\begin{eqnarray}
  S=S_{c}=\pi,
\end{eqnarray}
which corresponds to
\begin{eqnarray}
  r=r_{c}=1.
\end{eqnarray}
The temperature is $T_{c}=\frac{2}{\pi}$.\\
(3)when $\epsilon<24\pi$, $C_{\epsilon}>0$.

\begin{figure*}
\centerline{\includegraphics[scale=0.4]{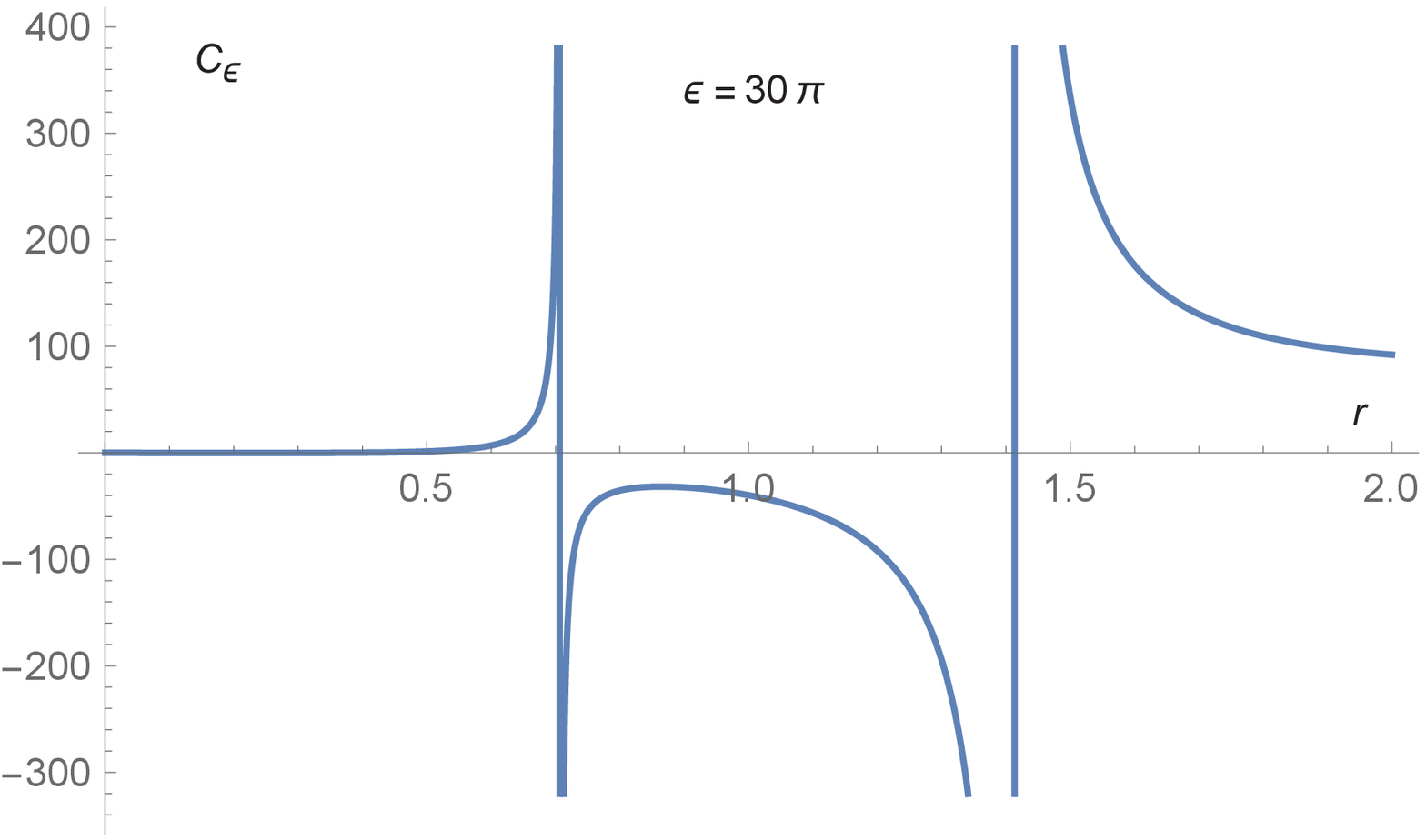}
\includegraphics[scale=0.4]{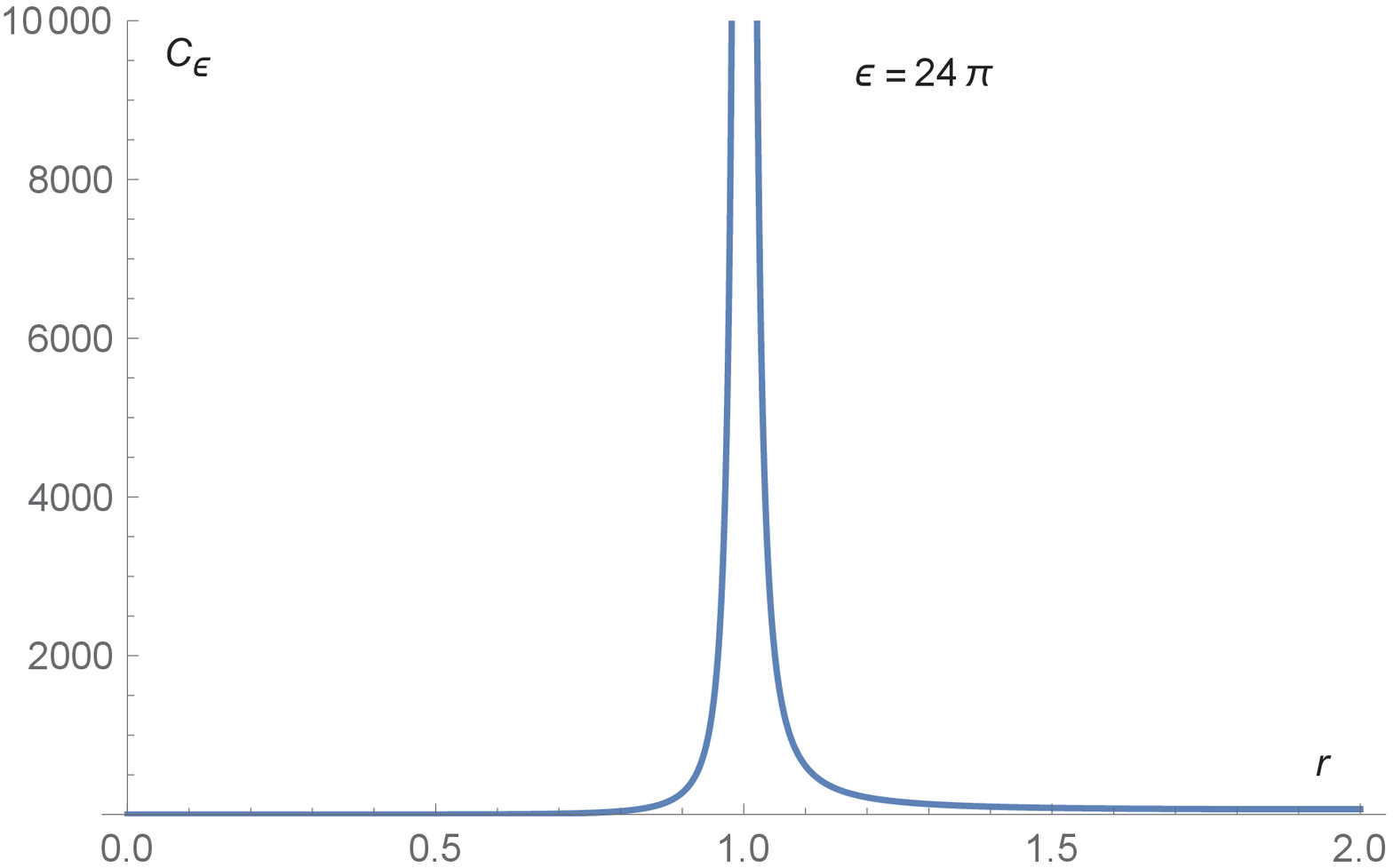}}
\centerline{\includegraphics[scale=0.4]{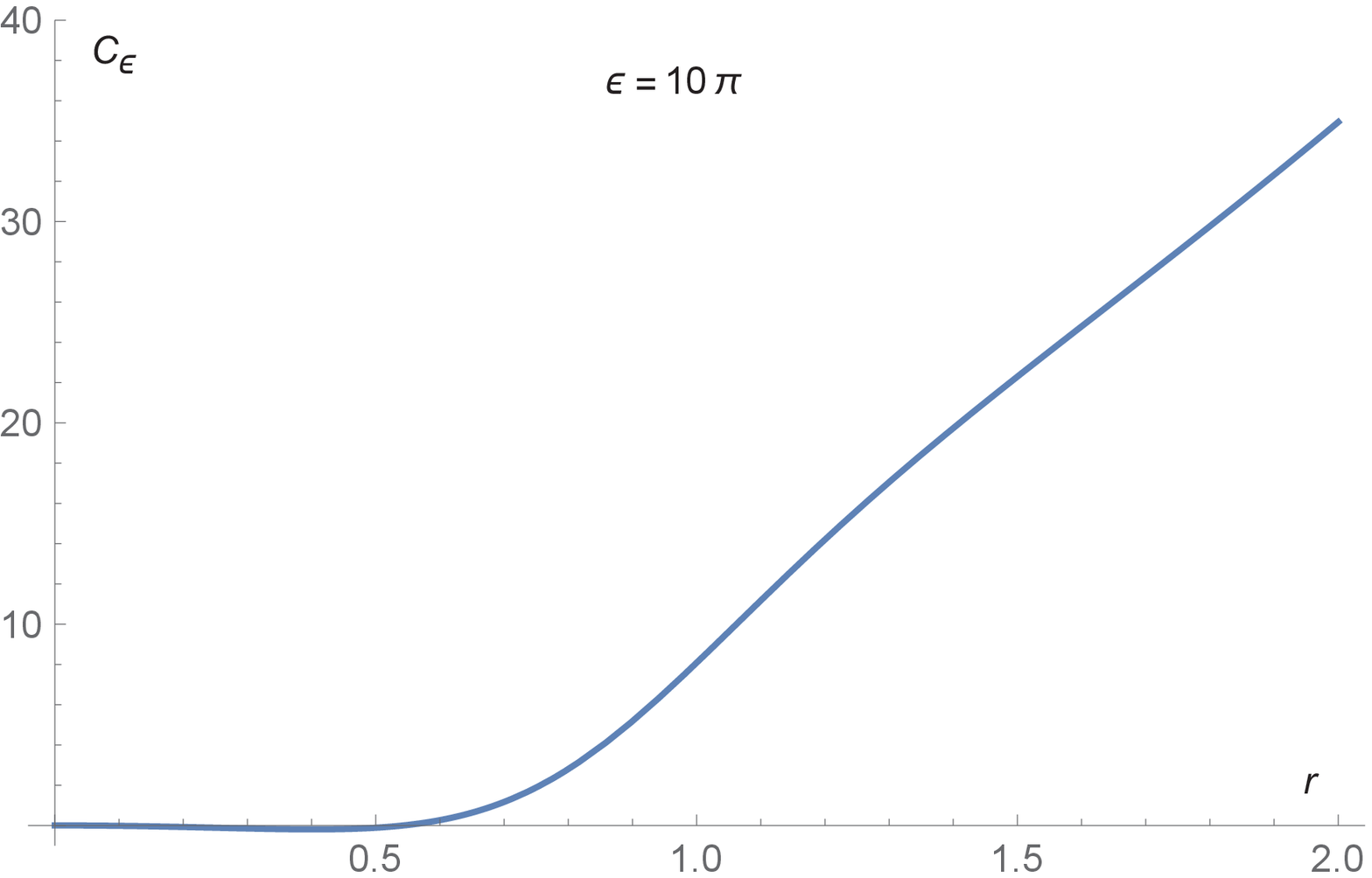}}
\caption{The specific heat $C_{\epsilon}$ vs. $r$ for $\epsilon=30\pi>\epsilon_{c}$ which has two divergent points,$\epsilon=24\pi=\epsilon_{c}$ which has only one divergent point and $\epsilon=10\pi<\epsilon_{c}$ which has no divergent point. }\label{specificheat}
\end{figure*}

Fig.\ref{specificheat} shows the behavior of specific heat for the cases $\epsilon>\epsilon_{c}$,$\epsilon=\epsilon_{c}$,$\epsilon<\epsilon_{c}$. For $\epsilon>\epsilon_{c}$, there are two divergent points on the specific heat curve,  they divide the region into three parts: the large radius region, the medium radius region and the small radius region. With positive specific heat, both the large radius region and the small radius region are thermodynamically stable.  While with negative specific heat, the medium radius region is unstable. So there is a phase transition take place between small black hole and large black hole. For $\epsilon=\epsilon_{c}$, the curve of specific heat  has only one divergent point and are always positive which denote that $\epsilon_{c}$ is the phase transition critical point. While for $\epsilon<\epsilon_{c}$ , the curve of specific heat has no divergent point and are always positive, which denote that the black holes are stable and no phase transition will take place.

\section{Van der Waals Like Behavior of Temperature  }
\label{tofrs}

It was shown in Ref.\cite{2012JHEP07033K} that when the cosmological constant is identified as thermodynamic pressure\cite{Kastor:2009wy}, $P-v$ graph exhibits van der Waals like behavior. Since these pioneering work, this universal property is discovered in various black holes\cite{Hendi:2017fxp,Hendi:2016vux,Hendi:2015soe,2013arXiv13053379S,Lan:2015bia,Ma:2017pap,Wei:2017icx,Bhattacharya:2017hfj,Hendi:2016usw,
Kuang:2016caz,Fernando:2016sps,Majhi:2016txt,Zeng:2016aly,Sadeghi:2016dvc,Zeng:2015wtt,Nguyen:2015wfa,Xu:2015rfa,
Nie:2015zia,Zhang:2014fsa,Belhaj:2014eha,Dehghani:2014caa,Mo:2013ela,Cai:2013qga,Chen:2013ce}.  Here, we find that for different topological charges, temperature of AdS black holes also possess the interesting van der Waals like property.

\begin{figure}
\centerline{
\includegraphics[scale=0.4]{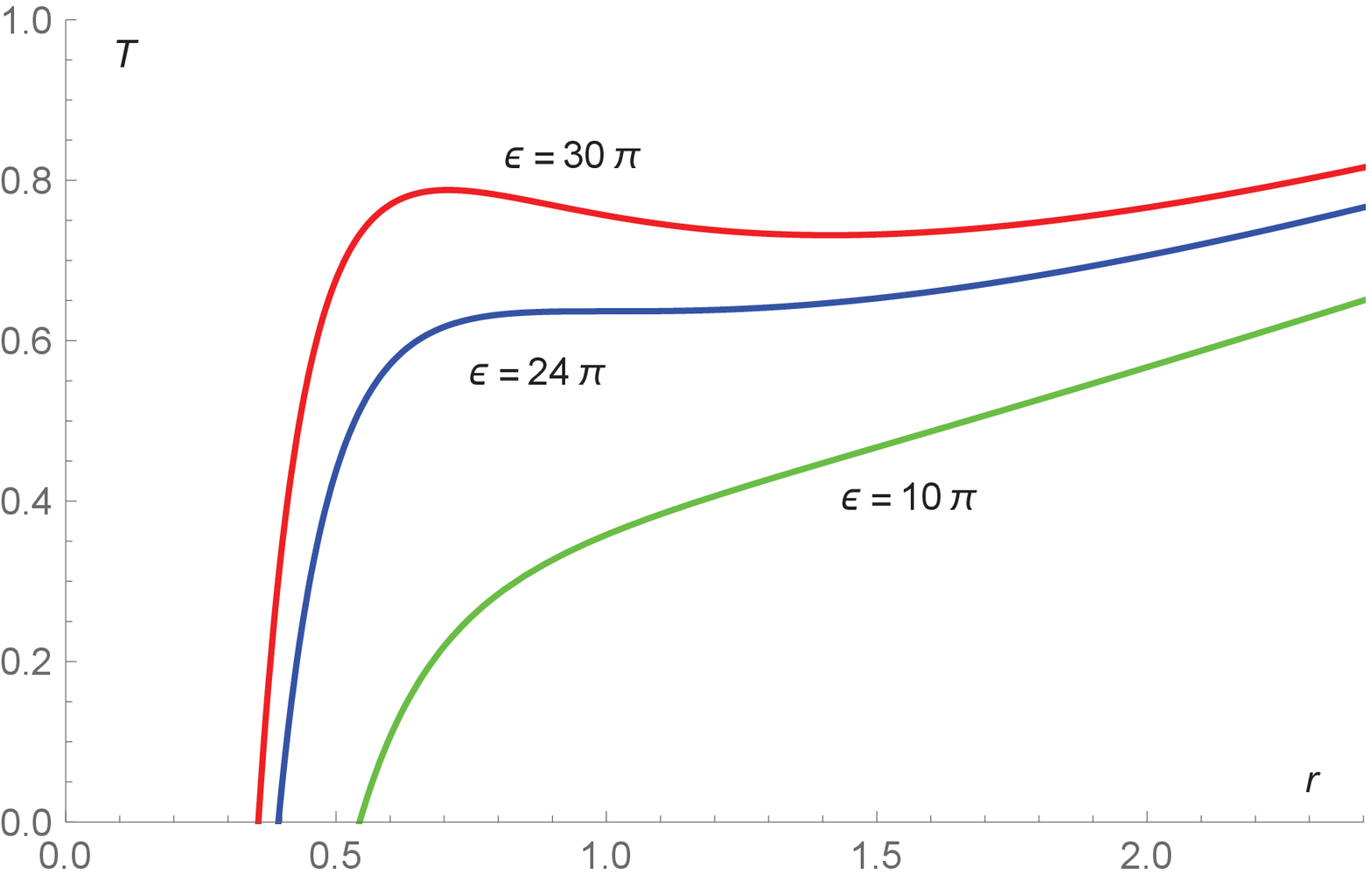}
\includegraphics[scale=0.40]{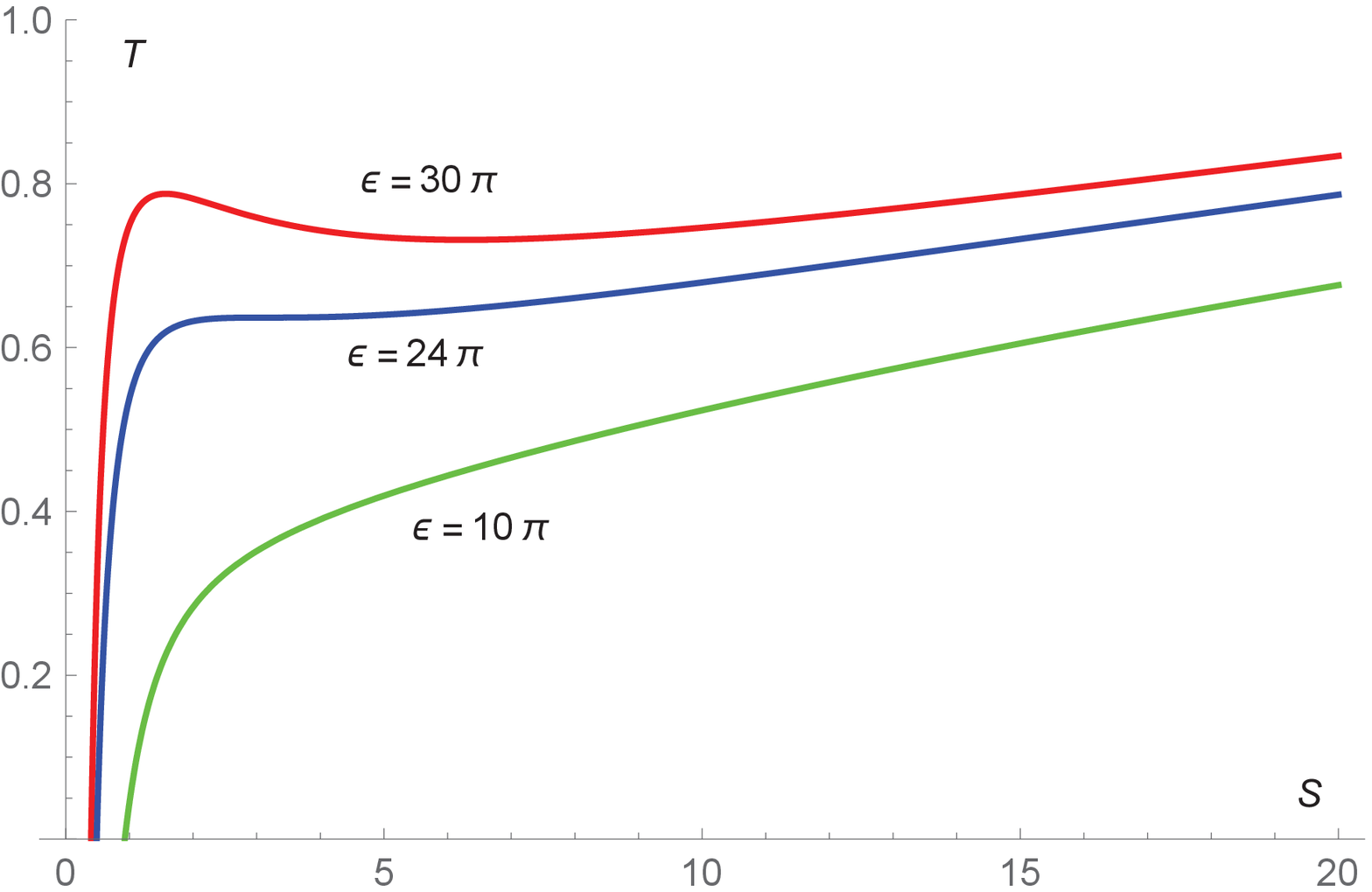}
}
\caption{$T$ vs. $r$ (left graph)  and  $T$ vs. $S$ (right graph) for different topological charge  $\epsilon$. There is an oscillating behavior when $\epsilon>\epsilon_{c}=24\pi$ for both $T(r)$ and $T(S)$ which is reminiscent of the van der Waals phase transition behavior. }\label{temrs}
\end{figure}

In $T-r$ curve, the possible critical point can be obtained by
\begin{eqnarray}
  &\,&(\frac{\partial T}{\partial r})_{\epsilon=\epsilon_{c},r=r_{c}}=0,\nonumber\\
  &\,&(\frac{\partial^{2} T}{\partial r^{2}})_{\epsilon=\epsilon_{c},r=r_{c}}=0.
\end{eqnarray}
Solving the above equations, one can obtain
\begin{eqnarray}
  \epsilon_{c}=24\pi,\,\,\,\,\,\,r_{c}=1,
\end{eqnarray}
which are exactly the same critical point we obtained by analysing the divergent behavior of specific heat.

In $T-S$ curve, the possible critical point can be obtained by
\begin{eqnarray}
  &\,&(\frac{\partial T}{\partial S})_{\epsilon=\epsilon_{c},S=S_{c}}=0,\nonumber\\
  &\,&(\frac{\partial^{2} T}{\partial S^{2}})_{\epsilon=\epsilon_{c},S=S_{c}}=0.
\end{eqnarray}
Solving the above equations, one can obtain
\begin{eqnarray}
  \epsilon_{c}=24\pi,\,\,\,\,\,\,S_{c}=\pi,
\end{eqnarray}
which are also exactly the same critical point we obtained by analysing the divergent behavior of specific heat.

Fig.\ref{temrs} shows the temperature behavior as a function of $r$ or $S$ for different values of topological charge $\epsilon$. When $\epsilon>\epsilon_{c}$, the curve can be divided into three branches. The slopes of the large radius branch and the small radius branch are both positive, while the slope of the medium radius branch is negative. When $\epsilon<\epsilon_{c}$, the temperature increases monotonically as increasing $r$ or $S$. This phenomenon is analogous to that of the van der Waals liquid-gas system.

From above, one can find that by analysing the specific heat curves, the $T-r$ curves and  the $T-S$ curves, the critical points are obtained and they are consistent with each other. In the above section, we have shown that both the large radius branch and the small radius branch are stable with positive specific heat, while the medium radius branch is unstable with negative specific heat. As argued in Ref.\cite{2013arXiv13053379S}, one can use the Maxwell equal area law to remove the unstable branch in $T-S$ curve with a bar vertical to the temperature axis $T=T^{*}$ and obtain the phase transition point $(T^{*},\epsilon)$. In the next section, we will use the Maxwell equal area law and analyse the free energy to determine the phase transition coexistence line.

\section{Maxwell Equal Area Law, Free Energy and Phase Diagram}
\label{maxfreephase}

\begin{figure}
\begin{center}
\includegraphics[scale=0.5]{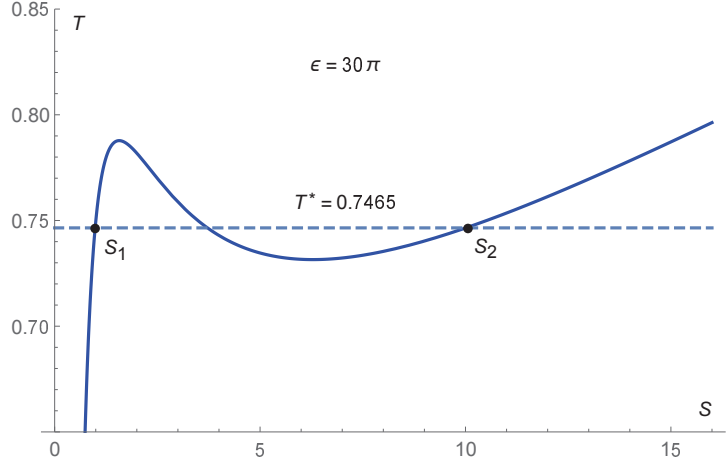}
\end{center}
\caption{$T$ vs. $S$ at $\epsilon=30\pi>\epsilon_{c}$. The dashed line $T=0.7465$ equally separate the oscillating part. According to the Maxwell's equal area law, the phase transition point is $(T=0.7465, \epsilon=30\pi).$ }\label{tsm}
\end{figure}

In Fig.\ref{tsm}, for fixed topological charge $\epsilon>\epsilon_{c}$, temperature $T(S,\epsilon)$ curve shows an oscillating behavior which denotes a phase transition. The oscillating part needs to be replaced by an isobar (denote as $T^{*}$) such that the areas above and below it are equal to each other. This treatment is called Maxwell's equal area law. In what follows, we will analytically determine this isobar $T^{*}$ for fixed $\epsilon$\cite{2013arXiv13053379S,Lan:2015bia}.

The Maxwell's equal area law is manifest as
\begin{eqnarray}
  T^{*}(S_{2}-S_{1})&=&\int_{S_{1}}^{S_{2}}T(S,\epsilon)dS\nonumber\\
  &=&\frac{1}{8\pi^{3/2}}(4S_{2}^{3/2}+\epsilon S_{2}^{1/2}+4\pi^{2}S_{2}^{-1/2}\nonumber\\
  &\,&-4S_{1}^{3/2}-\epsilon S_{1}^{1/2}+4\pi^{2}S_{1}^{-1/2}).
\end{eqnarray}
At points $(S_{1},T^{*})$,$(S_{2},T^{*})$, we have two equations
\begin{eqnarray}
  T^{*}&=&T(S_{1},\epsilon)=\frac{1}{16\pi^{3/2}}(12S_{1}^{1/2}+\epsilon S_{1}^{-1/2}-4\pi^{2}S_{1}^{-3/2}),\nonumber\\
  T^{*}&=&T(S_{2},\epsilon)=\frac{1}{16\pi^{3/2}}(12S_{2}^{1/2}+\epsilon S_{2}^{-1/2}-4\pi^{2}S_{2}^{-3/2}).
\end{eqnarray}
The above three equations can be solved as
\begin{eqnarray}
 && S_{1}=\frac{\epsilon-16\pi-\sqrt{(\epsilon-16\pi)^{2}}-64\pi^{2}}{8},\nonumber\\
  &&S_{2}=\frac{\epsilon-16\pi+\sqrt{(\epsilon-16\pi)^{2}}-64\pi^{2}}{8},\nonumber\\
 && T^{*}=\nonumber\\
  &&\frac{\epsilon^{2}-\epsilon\sqrt{\epsilon^{2}-32\pi\epsilon+192\pi^{2}}-28\pi\epsilon+12\pi\sqrt{\epsilon^{2}-32\pi\epsilon+192\pi^{2}}+160\pi^{2}}{\sqrt{2}\pi^{3/2}(\epsilon-16\pi-\sqrt{\epsilon^{2}-32\pi\epsilon+192\pi^{2}})^{3/2}}.
\end{eqnarray}
The last equation $T^{*}(\epsilon)$ is the phase transition curve we are looking for.

\begin{figure}
\begin{center}
\includegraphics[scale=0.5]{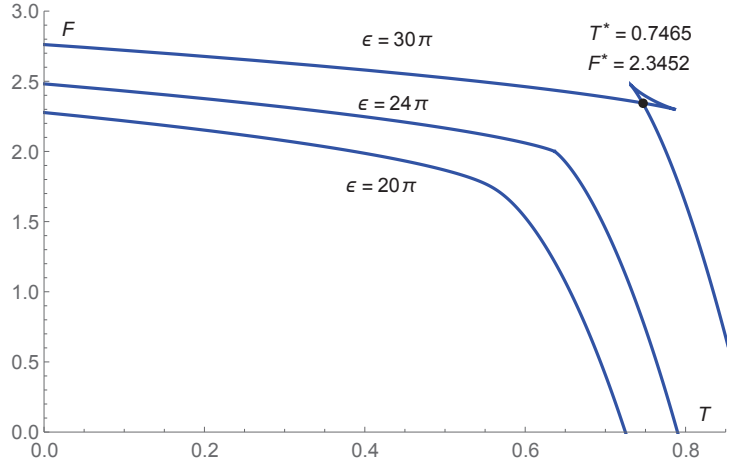}
\end{center}
\caption{$F$ vs. $T$ for different topological charge $\epsilon$. When $\epsilon>24\pi=\epsilon_{c}$, the curve shows a swallow tail behavior. }\label{fg}
\end{figure}

To double check the phase transition curve obtained by  the Maxwell's equal area law, we will probe the behavior of free energy, which is derived as
\begin{eqnarray}
  F=M-T S=\frac{12\pi^{2}+\epsilon S-4 S^{2}}{16\pi^{3/2}\sqrt{S}}.
\end{eqnarray}
Since temperature is also a function of $S$ and $\epsilon$, we can plot $F$ vs. $T$ in Fig.\ref{fg}. When $\epsilon>24\pi=\epsilon_{c}$, $F-T$ curve shows a swallow tail behavior which is reminiscent of $G-T$ curve for the van der Waals system. In this sense, the free energy here should be regarded as Gibbs free energy, and the inner energy $M$ should be regarded as enthalpy. Anyway, the cross point is determined by the equations below.

\begin{eqnarray}
  T^{*}&=&T(S_{1},\epsilon)=T(S_{2},\epsilon),\nonumber\\
  F^{*}&=&F(S_{1},\epsilon)=F(S_{2},\epsilon).
\end{eqnarray}
The right side equations can be rewritten as
\begin{eqnarray}
  &\,&\frac{1}{16\pi^{3/2}}(12S_{1}^{1/2}+\epsilon S_{1}^{-1/2}-4\pi^{2}S_{1}^{-3/2})\nonumber\\
  &=&\frac{1}{16\pi^{3/2}}(12S_{2}^{1/2}+\epsilon S_{2}^{-1/2}-4\pi^{2}S_{2}^{-3/2}),\nonumber\\
  &\,&\frac{1}{16\pi^{3/2}}(12\pi^{2}S_{1}^{-1/2}+\epsilon S_{1}^{1/2}-4 S_{1}^{3/2})\nonumber\\
  &=&\frac{1}{16\pi^{3/2}}(12\pi^{2}S_{2}^{-1/2}+\epsilon S_{2}^{1/2}-4 S_{2}^{3/2}).
\end{eqnarray}
These equations can be solved as
\begin{eqnarray}
 && S_{1}=\frac{\epsilon-16\pi-\sqrt{(\epsilon-16\pi)^{2}}-64\pi^{2}}{8},\nonumber\\
  &&S_{2}=\frac{\epsilon-16\pi+\sqrt{(\epsilon-16\pi)^{2}}-64\pi^{2}}{8},\nonumber\\
 && F^{*}=\frac{\epsilon-\sqrt{\epsilon^{2}-32\pi\epsilon+192\pi^{2}}-8\pi}{2\sqrt{2\pi}\sqrt{}\epsilon-16\pi-\sqrt{\epsilon^{2}-32\pi\epsilon+192\pi^{2}}},\nonumber\\
 && T^{*}=\nonumber\\
 &&\frac{\epsilon^{2}-\epsilon\sqrt{\epsilon^{2}-32\pi\epsilon+192\pi^{2}}-28\pi\epsilon+12\pi\sqrt{\epsilon^{2}-32\pi\epsilon+192\pi^{2}}+160\pi^{2}}{\sqrt{2}\pi^{3/2}(\epsilon-16\pi-\sqrt{\epsilon^{2}-32\pi\epsilon+192\pi^{2}})^{3/2}}.
\end{eqnarray}
They are consistent with the results obtained by the  Maxwell's equal area law.

\begin{figure}
\begin{center}
\includegraphics[scale=0.5]{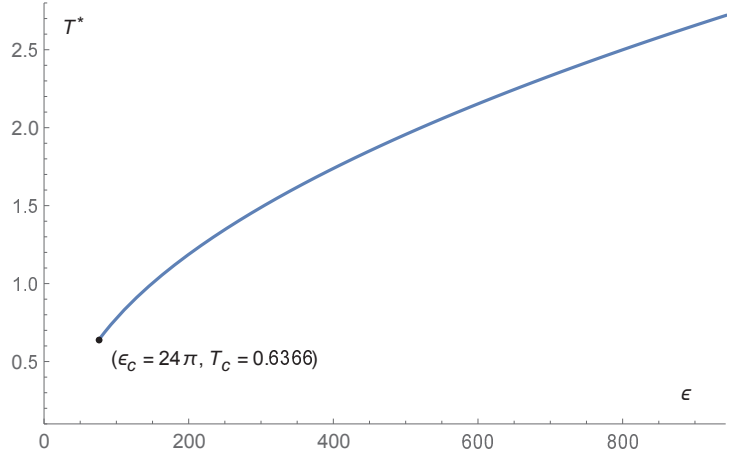}
\end{center}
\caption{The phase transition coexistence line $T^{*}-\epsilon$ for fixed electric charge $Q=1$ and AdS radius $l=1$.}\label{phasete}
\end{figure}

Finally, we can show the phase transition coexistence line in Fig.\ref{phasete} for fixed electric charge $Q$ and AdS radius $l$. This kind of phase transition is special, as the critical point is at the small value of $(T^{*},\epsilon)$ in phase diagram.

\section{Conclusion}
\label{con}

In this paper, the phase transition phenomenon of Reissner-Nordstr$\ddot{o}$m AdS black holes relating to the topological charge $\epsilon$ in canonical ensemble in 4 dimensional space-time are studied. As we are interested in the effects of the topological charge, so the electric charge is fixed $Q=1$. Firstly, the black hole's specific heat $C_{\epsilon}$ is calculated and the corresponding divergence solutions are derived. The two solutions merge into one denoting the critical point where $\epsilon_{c}=24\pi, r_{c}=1$. When $\epsilon>\epsilon_{c}$, the curve of specific heat has two divergent points and is divided into three regions. The specific heat are positive for both the large radius region and the small radius region which are thermodynamically stable, while it is negative for the medium radius region which is unstable. When $\epsilon<\epsilon_{c}$, the specific heat is always positive implying the black holes are stable and no phase transition will take place.

Secondly, the behavior of temperature in both the $T-r$ graph and $T-S$ graph are studied. They exhibit the interesting van de Waals gas-liquid system's behavior. The critical points correspond to the inflection points of $T-r$ curve and $T-S$ curve, and they are consistent with that derived from the specific heat analysis. When $\epsilon>\epsilon_{c}$, the curves can be divided into three regions. The slope of the large radius regions and the small radius regions are positive while those of the medium radius region are negative.  When $\epsilon<\epsilon_{c}$, the temperature increase monotonically.

Thirdly, a van der Waals system's swallow tail behavior is observed when $\epsilon>\epsilon_{c}$ in the $F-T$ graph. What's more, by using the Maxwell's equal area law and analysing the free energy, the analytic phase transition coexistence lines are obtained, and they are consistent with each other.

From the above detailed study, one can find that this van der Waals like system exhibits phase transition of special property. The phase transition take place at large topological charge $\epsilon>\epsilon_{c}$ and high temperature which can be clearly seen from the phase transition coexistence line in Fig.\ref{phasete}. Whether this phase transition property is universal in other gravity theories ( such as the Lovelock, Gauss-Bonnet theory ) and different dimensional space-time is unknown.

There are some other interesting topics that are worth investigating, such as, the holographic duality in the field theory of this kind of phase transition; the cases of space-time dimension $d>4$, as the topological charge's conjugate potential $\omega=\frac{1}{8\pi}K^{\frac{4-d}{2}}r^{d-3}$ decayed to $\frac{1}{8\pi r}$ which is irrelevant to $K$ at $d=4$ dimensional space-time.

 \section*{Acknowledgements}
 This research is in part supported by National Natural Science Foundation of China (Grant Nos.11605082,11747017), Natural Science Foundation of Guangdong Province, China (Grant Nos.2016A030310363, 2016A030307051, 2015A030313789) and Department of Education of Guangdong Province, China
 (Grant Nos.2017KQNCX124, 2017KZDXM056).


\begin{thebibliography}{99}




\bibitem{Chen:2010ay}
     Y. X. Chen, J. L. Li,
     First law of thermodynamics on holographic screens in entropic force frame,
      Phys. Lett. B
      700
      (2011)
     380-384

\bibitem{Brown:1992br}
     J.D. Brown, J. York, W. James,
     Quasilocal energy and conserved charges derived from the
                        gravitational action,
    Phys. Rev.D
      47
      (1993)
      1407-1419


\bibitem{Brown:1992bq}
    J. D. Brown, J. York, W. James ,
   The Microcanonical functional integral. 1. The
                        Gravitational field,
      Phys. Rev.D 47
      (1993)
     1420-1431

\bibitem{Bredberg:2011jq}
    I. Bredberg, C. Keeler, V. Lysov, A. Strominger,
     From Navier-Stokes To Einstein,
      JHEP 07
      (2012)146

\bibitem{Tian:2014goa}
     Y, Tian,X. N. Wu, H. B. Zhang,
     Holographic Entropy Production,
    JHEP
     10
      (2014)
      170

\bibitem{Tian2018hlw}
Y. Tian, \emph{The last (lost) charge of a black hole}, \emph{arXiv:1804.00249,gr-qc} (2018).

\bibitem{Tian:2010gn}
      Y. Tian, X. N. Wu,
    Dynamics of Gravity as Thermodynamics on the Spherical
                        Holographic Screen,
     Phys. Rev.D,
     83
      (2011)
     021501

\bibitem{Wald:1993nt}
R. M. Wald, Black hole entropy is the Noether charge, Phys. Rev. D 48 (1993) 3427-3431

\bibitem{Padmanabhan:2010xh}
     T. Padmanabhan,
    Surface Density of Spacetime Degrees of Freedom from
                        Equipartition Law in theories of Gravity,
   Phys. Rev.D 81
      (2010)
      124040

\bibitem{Mo:2016apo}
     J. X. Mo, G. Q. Li, Y. C. Wu,
     A consistent and unified picture for critical phenomena
                        of $f(R)$ AdS black holes,
     JCAP
      1604
      (2016)
     045



\bibitem{2012JHEP07033K}
 D. {Kubiz{\v n}{\'a}k}, R. B. Mann,
   P - V criticality of charged AdS black holes,
  JHEP 7
     (2012)
  33

\bibitem{Kastor:2009wy}
      D. Kastor, S. Ray, J. Traschen,
     Enthalpy and the Mechanics of AdS Black Holes,
    Class. Quant. Grav.
     26
     (2009)
     195011

\bibitem{Hendi:2017fxp}
    S. H. Hendi, R. B. Mann, S. Panahiyan, B. Eslam Panah,
    Van der Waals like behavior of topological AdS black
                        holes in massive gravity,
    Phys. Rev.D
    95
     (2017)
     021501

\bibitem{Hendi:2016vux}
S. H. Hendi, B. Eslam Panah, S.Panahiyan,
     Topological charged black holes in massive gravity's
                        rainbow and their thermodynamical analysis through various
                        approaches,
     Phys. Lett.B
      769
      (2017)
      191-201

\bibitem{Hendi:2015soe}
    S. H. Hendi, S. Panahiyan, B. Eslam Panah,
     Extended phase space of Black Holes in Lovelock gravity
                        with nonlinear electrodynamics,
    PTEP 10
      (2015)
      103E01

\bibitem{2013arXiv13053379S}
 E. Spallucci A. Smailagic,
    Maxwell's equal area law for charged Anti-deSitter black holes,
  Phys. Let. B  723(2013)436

\bibitem{Lan:2015bia}
      S. Q. Lan, J. X. Mo, W. B. Liu,
     A note on Maxwell's equal area law for black hole phase
                        transition,
   Eur. Phys. J. C
     75
      (2015)
      419


\bibitem{Ma:2017pap}
     M. S. Ma, R. H. Wang,
     Peculiar P-V criticality of topological
                        Horava-Lifshitz black holes,
     Phys. Rev. D
     96
     (2017)
     024052

\bibitem{Wei:2017icx}
     S. W. Wei, B. Liang, Y. X. Liu,
    Critical phenomena and chemical potential of a charged
                        AdS black hole,
    Phys. Rev. D
      96
      (2017)
     124018

\bibitem{Bhattacharya:2017hfj}
   K. Bhattacharya, B.R. Majhi,
     Thermogeometric description of the van der Waals like
                        phase transition in AdS black holes,
   Phys. Rev. D
    95
      (2017)
     104024

\bibitem{Hendi:2016usw}
     S. H. Hendi, B. Eslam Panah,S. Panahiyan, M.S. Talezadeh,
     Geometrical thermodynamics and P-V criticality of black
                        holes with power-law Maxwell field,
      Eur. Phys. J. C
      77
      (2017)
     133


\bibitem{Kuang:2016caz}
   X. M. Kuang, O. Miskovic,
   Thermal phase transitions of dimensionally continued AdS
                        black holes,
     Phys. Rev. D
     95
      (2017)
      046009

\bibitem{Fernando:2016sps}
     S. Fernando,
     P-V criticality in AdS black holes of massive gravity,
   Phys. Rev. D
    94
      (2016)
     124049

\bibitem{Majhi:2016txt}
     B. R. Majhi,S.Samanta,
     P-V criticality of AdS black holes in a general
                        framework,
     Phys. Lett. B
     773
     (2017)
     203-207

\bibitem{Zeng:2016aly}
   S. He, L. F.  Li, X. X. Zeng,
    Holographic Van der Waals-like phase transition in the
                        Gauss¨CBonnet gravity,
     Nucl. Phys.B
    915
      (2017)
    243-261

\bibitem{Sadeghi:2016dvc}
     J. Sadeghi, B. Pourhassan, M. Rostami,
     P-V criticality of logarithm-corrected dyonic charged
                        AdS black holes,
    Phys. Rev.D
      94
     (2016)
     064006

\bibitem{Zeng:2015wtt}
     X. X. Zeng, L. F. Li,
     Van der Waals phase transition in the framework of
                        holography,
     Phys. Lett.B
      764
     (2017)
      100-108

\bibitem{Nguyen:2015wfa}
     P. H. Nguyen,
     An equal area law for holographic entanglement entropy
                        of the AdS-RN black hole,
      JHEP
      12
      (2015)
      139

\bibitem{Xu:2015rfa}
    J. F. Xu, L. M. Cao, Y. P. Hu,
     P-V criticality in the extended phase space of black
                        holes in massive gravity,
     Phys. Rev. D,
    91
     (2015)
     124033

\bibitem{Nie:2015zia}
     Z. Y. Nie, H. Zeng,
      P-T phase diagram of a holographic s+p model from
                        Gauss-Bonnet gravity,
     JHEP 10
      (2015)047


\bibitem{Zhang:2014fsa}
     L. C. Zhang, H. H. Zhao, R.Zhao, M. S. Ma,
    Equal Area Laws and Latent Heat for $d$-Dimensional
                        RN-AdS Black Hole,
     Adv. High Energy Phys.
     2014
     (2014)816728


\bibitem{Belhaj:2014eha}
    A. Belhaj, M. Chabab, H. El moumni, H. K. Masmar,
                       M. B.Sedra,
     Maxwell`s equal-area law for Gauss-Bonnet-Anti-de Sitter
                        black holes,
    Eur. Phys. J. C
     75
      (2015)71


\bibitem{Dehghani:2014caa}
    M. H. Dehghani, S. Kamrani, A.Sheykhi,
     $P-V$ criticality of charged dilatonic black holes,
  Phys. Rev.D
    90
    (2014)104020

\bibitem{Mo:2013ela}
     J. X. Mo, W. B. Liu,
   Ehrenfest scheme for P-V criticality in the extended
                        phase space of black holes,
   Phys. Lett.B 727(2013)336-339


\bibitem{Cai:2013qga}
   R. G. Cai, L. M. Cao, L. Li, R. Q. Yang,
    P-V criticality in the extended phase space of
                        Gauss-Bonnet black holes in AdS space,
     JHEP
      09
     (2013)005


\bibitem{Chen:2013ce}
   S. B. Chen, X. F. Liu, C. Q. Liu, J.L. Jing,
     $P-V$ criticality of AdS black hole in $f(R)$ gravity,
      Chin. Phys. Lett. 30(2013)060401




\end{thebibliography}
\end{document}